# The Impact of Project Management in Virtual Environment: A Software Industry Perspective*

N. B. J. Gamage
Faculty of Management Studies & Commerce
University of Sri Jayawardhanapura
Gangodawila,
Nugegoda, Sri Lanka
bashini.jeewanthi@gmail.com

*Abstract*—**Virtual team in a project within an organization could achieve optimize project performance by acquiring appropriate human resources, coordination, communication and regular performance evaluation. According to the literature many ICT tools will collaborate to manage virtual teams, but still most of the projects lead to failure in the software industry. Aim of this research is to discover the most affected factors for virtual project human resource management.**

## I. INTRODUCTION

A virtual team is a group of individuals who work across time, space and organizational boundaries with links strengthened by communication technology (Lipnack, 2000). Virtual team concept is very common in the software industry. Most of the occasions clients and software development team in different locations in the world. Sometimes there are multiple virtual development teams located in different places. Their subject experts live in different places in the world. So that project manager needs to hire team from all over the world locations.

As Cristian and Adela's study most of virtual teams are ineffective as a percentage 25% , lack of team performance as a percentage 27%. Another issue is individual performance lower of team members as a percentage 17%. But another study prove that 18% virtual teams are success in action. Cristian and Alde emphasis three main reasons affected to the virtual team in trouble. They were, "Lack of face-to-face contact with team members (46% of responses), Lack of Resources (37%) and Time zone differences hindering the ability to collaborate (29%)"( Cristian and Aldea, 2014, p.380).

Lilian(2013) argues that manage virtual team should need good and strong communication plan and control among between all team members. Managing human resources in the virtual team is also big responsibility of team project manager or e-leader. Papadopoulos (2015) state that most of the software project managers use Agile methodology as project management tool in virtual environment. Agile project management is process conducting with active support of all stakeholders of the project. Main advantage of this method Agile can easily apply to the virtual teams as well. In a virtual environment team members are located in different places in the world. Basically "tracking the team members performances, provide them to sufficient feed backs, resolving their burning issues and optimizing the project performance"(Inc Project Management Institute, 2008, p.215).

Therefore, this research investigates Major factors that could influence effective Project Human in virtual environment of Sri Lankan Software Industry? Lilian(2013) argues that virtual team human resource management is challengeable task for practicing project manager. Most of the projects fails because resource inefficiency and late to discover resource skill gap. Gain maximum output from the project team in a virtual environment affected to project managers' performances as well as success of the project. Furthermore virtual team human resource management has challengeable issues, so to find solution for these issues I suppose to conduct this research. This study will benefit by the project managers and other all stakeholders who handle their projects in virtual environment.

The objectives of this study are as follows:

· To identify what are practicing project managers facing issues.

· To identify factors influence project human resource management in Virtual environment.

In addition, the scope of this research encompasses how project managers' handle their project team members in a virtual environment. How do they use Agile project management methodology effectively and efficient way to handle the problems while project running? What are the major factors affected to the project human resource management in a virtual team?

## II. THEORETICAL BACKGROUND

*A. Effective project Management*

*1) Virtual team Management*

Virtual team is group of members who working at different locations in the world and they communicate and collaborate work by using ICT tools. Apart from the communication between team members "knowledge sharing, transfer, acquisition, integration and archive"( Aldea et al., 2012, p. 650) doing through ICT tools. "At the same time, Virtual Team have the potential to achieve further gains in processes and provide high quality solutions by meeting, gathering people with different knowledge, expertise"( Aldea et al., 2012, p. 650). According to the Cosmina(2015) findings these are the issues affected to the performance of the Virtual Team, such as Elevating goal, structure, Members' competencies(commitment and Trust), Building a collaborative climate, Standard of excellence and external support and Leadership.

*B. Agile project management*

Papadopoulos(2015)argues that most used and effective approach to virtual team managing methodology is Agile methodology. Agile approaches are used to help businesses respond to unpredictable. "They provide opportunities to assess the state of play and the direction of the project at different points over time" (Amorim and Sousa, 2014, p. 871). "Having cross-linked, self-organized teams and a flat organizational structure allows agile teams to closely collaborate without needless complications"(Papadopoulos, 2015, p. 456). Agile methodology has frequent releases and its move to with flexible changes of stakeholders or clients. It always welcome to changes.

*C. Project Human Resource Management*

Human Resource management process contains main elements such as Human Resource Planning, Acquire Project Team, Develop Project Team and Manage Project Team (Inc Project Management Institute, 2008, p.215). In HR planning, project manager should identify and record as document project roles, responsibilities, reportinghierarchy. According to the project scope project manager should estimate size of the project team and from where they want to hire human resources. Therefore he need to identifying and documenting the project roles and responsibilities as well as reporting relationships."Resolving attitudes and conflicts, encouraging individuals for teamwork and identify business units"(Tohidi, 2010, p.926)

Acquire appropriate Project Team is very important. Because inappropriate human resources "causes obvious and hidden problems"(Tohidi, 2010, p.926). Therefore project manager should give high attention is to the personality characteristics of individual. "In IT organizations, hiring qualified people is very important job meaning individuals who self-reliable and responsible are suitable for IT projects." (Tohidi, 2010, p.926)

Develop Project Team is highly responsible task for project manager(Inc Project Management Institute, 2008, p.215). Members of a virtual team coordinate their work, predominantly with electronic information and communication technologies (e-mail, video-conferencing, telephone, etc.) (Hertel et al., 2005, p. 69) Project manager needs to improving the competencies of the team members and interaction with them to enhance project performance. "For that he can use One-on-one meetings, Modeling the 'AS-IS' Business process and Modeling the 'TO-BE' Business process" (Cristian and Aldea, 2014, p.383). Slow or delayed feedback due to communication media has negative impact on global project team performance. Stawnicza(2014) argues that usage of asynchronous communication tools, such as e-mail, discussion boards, shared documents, web logs, etc. for solving urgent issues, the lack of immediate response can delay the decision making process. The communications medium with the highest level of richness is face-to-face communication, followed by video conferencing, phone, and chat respectively.

"The lowest richness level is represented by e-mail, text messaging and written documents" (Weimann et al.,2010, p. 187). Sharing knowledge, information and other relevant details is most important fact of the team. Most engineering projects, suits of tools are also used "(e.g. Google Apps for Education, code hosting applications - https://bitbucket.org/, https://github.com/, file sharing applications - https://www.dropbox.com/, https://drive.google.com)" (Dascalu et al., 2015, p.104).

Overall responsibility is to manage Project Team. As essential steps of the project management we can take Tracking team member performance, providing feedback, resolving issues, and coordinating changes to enhance project performance. There are many software tools available for measure project performance in different angles. "BCT (Basic Communication Technologies) are especially used for high-performance projects, EST (Enterprise Software Technologies), e.g., ERP systems and project management software packages, are desirable for projects where the environment is well structured, and the GCT (Group Collaboration Technologies) must be given special weight for projects where the environment is less structured, uncertain and volatile" (Pellerin et al., 2013, p. 860). These above software tools helps to optimize real time communication between team members of the virtual team. Any Meeting (Known as the Freeboard and it has all you need to make an online appointment: send invitations via e-mail create custom registration forms, promote your meetings on Twitter and Facebook, PC and MAC compatibility, allows up to 200 participants and unlimited number of meetings(http://www.anymeeting.com/)), Caltech's EVO.(It is the successor of videoconferencing systems Virtual Room Video Conferencing and is a tool produced by Caltech. There is no limit on how many people can participate. "Client can connect to any server EVO called Panda.), Live Meeting (It can streamlined console design, rich multimedia options, webcam, audio controls, schedule meetings and invite participants, live Q&A (audience members can ask questions and get answers without interrupting the presenter)) and Skype(Millions of individuals use Skype for audio and video free calls, to send instant messages and files or for group meetings with other Skype users. It can Skype-to-Skype, call landlines and mobiles, conference calling, Skype to go, online number, voicemail, call forwarding, caller ID and call transfer)"( Aldea and Olariu, 2014, p. 334).

## III. CONCEPTUAL FRAMEWORK

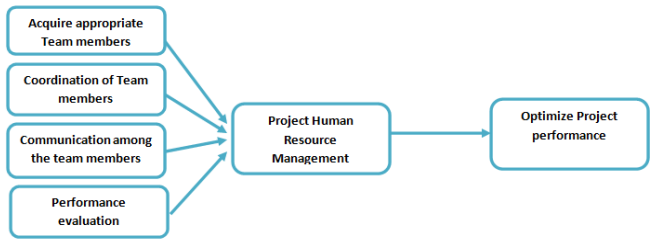

**Figure 1: Conceptual Framework**

The following hypotheses were developed based on the above conceptual framework.

*Acquire appropriate Team members:*
$H_{01}$:There is no relationship between acquiring appropriate Team members and project performance.
$H_{a1}$:There is relationship between acquiring appropriate Team members and project performance.

*Coordination of Team members:*
$H_{02}$:There is no relationship between Coordination of Team members and project performance.
$H_{a2}$:There is relationship between Coordination of Team members and project performance.

*Communication among the team members:*
$H_{03}$:There is no relationship between Communication among the team members and project performance.
$H_{a3}$:There is relationship between Communication among the team members and project performance.

*Performance evaluation:*
$H_{04}$:There is no relationship between Performance evaluation and project performance.
$H_{a4}$:There is relationship between Performance evaluation and project performance.

## IV. METHODOLOGY

### A. Research Philosophy

According to the literature review, most of the researchers did studies human resource management of software projects, but little work has been considered for virtual team handling. However as of this aspect I select analyze key factors which impact to the Virtual team Human Resource management by using deductive method. For deductive approaches the emphases is generally on causality (Gabriel, 2015). This proposal aims to investigate how traditional human resource management applies to the virtual team in a given project?

### B. Research Strategy

According to my literature review I identified above variables for affecting to the project performance through the project human resource management. Qualitative analysis use for interpret some variables. I need to identify how do the aforementioned factors affect to the employees' performance of a project, which is managing in a virtual environment. Therefore I choose Quantitative method for conduct this research.

To investigate the issues surrounding acquiring appropriate human resources, Coordination, communication and regular performance evaluation, a critical approach is used from a social critique and a survey study is carried out to collect data. Respondents are targeted as individual team members in the context of organizational context.

" There are several research methodology approaches available in Information Systems literature"(Arachchilage, 2012; Arachchilage and Love, 2013; 2014; Arachchilage, 2015; Arachchilage and Martin, 2015; Arachchilage and Martin, 2013; Arachchilage, Namiluku and Martin, 2013). For example, qualitative, quantitative and mixed method approach. According to the research methodology, qualitative approach could have been used in this research study. However, quantitative approach is selected because," it offers the flexibility to represent the population in general of users within organizations and also widely penetrated approach in IS" (Arachchilage and Love, 2013; Arachchilage and Love, 2014) Furthermore, quantitative research approach can aid to use tables and charts to visualize the data, use appropriate means to describe it, and chose some methods to examine trend and relationship within it using statistical techniques.

Interpretive studies do not test a hypothesis, as compared to positivist research (Arachchilage et al., 2013; Arachchilage and Martin, 2015). Also, if the research is based on a survey or experiments, it is more likely to be based on positivism. So, this proposed research approach would be positivist research because it is based on a survey and testing hypothesis derived from a research model. Therefore, positivist research could be more appropriate because the focus of this proposed research is to identify the factors that influence to acquiring appropriate human resources, Coordination, communication and regular performance evaluation.

The idea of a survey is used to obtain same sorts of data from a large group of people in a standard and systematic way (Arachchilage, 2012). On the other hand, it may not consume respondents' time unnecessarily.

Perhaps the greatest benefits offered by a survey study from respondent's viewpoint would be the freedom to choose their best opinion from a given set of options, which is also more easy process for them instead of having an interview (Arachchilage et al, 2015). Furthermore, overall survey study could help respondents to get an idea of the research field.

### C. Population

Project managers and Team Leads of Software development companies who manage virtual teams in the Sri Lanka.

### D. Sample

I plan to conduct this research for 10 teams software development virtual team with 7-10 members, who work different locations in the world.

*E. Operationalization of variable*

Here I mentioned I identified independent variables and measurement using a five-point scale Likert at 1 = 'Strongly disagree' and 5 = 'Strongly agree'.

| Variable | Measurement |
|---|---|
| Acquire appropriate Team members | Likert Scale |
| Coordination of Team members | Likert Scale |
| Communication among the team members | Likert Scale |
| Performance evaluation | Likert Scale |

*F. Proposed statistical method*

Both conceptual and procedural knowledge were addressed by a total 10 question items which were evaluated based on participants level of knowledge. For analysis of quantitative data I suppose to use multiple regression method and correlation analysis

## V. CONCLUSION

An organization with virtual team and project could achieve optimize project performance by acquiring appropriate human resources, Coordination, communication and regular performance evaluation. According to the literature many ICT tools collaborate to manage virtual teams, but still most of the projects fails in the software industry. Aim of this research is to discover the factors that affect virtual project management in a software organization.